# Free space strange and unipolar EM pulses: yes or no?

Nikolay L. Popov*, Alexander V. Vinogradov

P.N. Lebedev Physical Institute, Leninsky Prospekt 53, Moscow, Russia 119991
* Correspondence: popovnl@sci.lebedev.ru

**Abstract:** E.G. Bessonov suggested the time integrated strength of electric field $\int_{-\infty}^{\infty} \boldsymbol{E}(r,t)dt = S_E(r)$ as a parameter to classify electromagnetic (EM) waves. Since then, this parameter has been studied and used in many works on microwave and laser physics, especially when it comes to unipolar, bipolar and few cycle EM pulses. In this paper, it is shown that $S_E(r) = 0$ is an identity for a wide class of free space pulses of finite total energy. This property can be useful in various applications of few cycle radiation and as a benchmark in EM and QED computations.

## 1. Introduction

The development and widespread use of few cycle lasers [1] motivated the study of little-known and new properties of free electromagnetic (EM) waves, or more precisely, pulses [2–4]. By pulse we mean the solution of Maxwell's equations in the absence of sources, corresponding to a finite energy. Obviously, such solutions should describe real pulses propagating in free space. The question posed in the title of the article also refers to real pulses. To answer it, it is necessary to turn to the full set of electromagnetic wave equations and add the only requirement: the finiteness of the energy of the electromagnetic field. Thus, monochromatic and plane waves, are out of our consideration, since their energy is infinite; Gaussian beams too - they, moreover, do not satisfy Maxwell's equations, as well as some other commonly used models.

Let's now go back to few cycle laser pulses of finite energy. Their distinctive property, which is clearly manifested and often discussed, is space time couplings (STC). This implies that the spatial configuration of the field of the propagating pulse is constantly changing with time, while the time shape and the spectrum of the pulse change from point to point. For example, it is well known that when focusing, the shape and spectrum of the compressed pulse differ from the incident one, and it is not a matter of duration, but of the number of periods of the field. Any pulse of finite energy has the STC property. However, when the number of periods of the field N is large, these effects are weak and are not given importance. It is enough to characterize the structure of the pulse and its interaction with matter in terms of the average frequency $\omega$ and line width $\delta\omega$. If the pulse is short-period and N approaches 1, the values $\omega$ and $\delta\omega$ become close. In this case, the result of the impact of a pulse on a substance essentially depends on its shape, and not on these average characteristics of the spectrum [5].

For a consistent description of STC effects, it is necessary to refer to models of finite energy pulses with nonseparable dependence on space and temporal coordinates. This can be achieved in exact solutions of Maxwell's equations, as well as in other rigorous methods. It should be noted here that, as a rule, finding exact solutions and studying their properties is also a difficult task [2,6-10]. On a qualitative level, we can say that STC is expressed in the extreme variability of any real electromagnetic pulse in time and space [11].

This work is devoted to another, in a sense, opposite property of electromagnetic pulses, which, like STC, is also important in ultrafast optics. This is due to the once rarely used characteristic of the EM field

$$S_E(r) = \int_{-\infty}^{\infty} E(r,t)dt, \quad (1)$$

which in the era of few cycle pulses attracted increased attention. This value first appeared in the work [12] by E.G. Bessonov, who studied radiation of particles in accelerators. To describe the radiation field $E(r,t)$ of a system of charged particles, he introduced parameter (1) and proposed to use it to classify electromagnetic waves, calling waves with

$$S_E(r) \neq 0 \quad (2)$$

strange. At the same time, as Bessonov showed, the radiation field of sources of a sufficiently wide class, namely, any system of charges performing a finite motion, satisfies the relation:

$$S_E(r) = 0. \quad (3)$$

He called EM waves, for which (3) is satisfied for all $r$, usual. Thus, in the classification proposed by Bessonov, usual waves, are necessarily bipolar. At the same time, strange waves, according to (2), can be both bipolar and unipolar (single sign). Their source can be charges that perform infinite motion, including bremsstrahlung, Compton scattering, the radiation of charged particles in bending magnets, the radiation of cosmic rays in the magnetic field of the Earth, radiation of electrons reflected from the surface of crystals, etc. [12]. The work of Bessonov received a noticeable response in accelerator and microwave communities. Since then, several theoretical and experimental papers on e-beam and other sources of bipolar and unipolar waves have been published (see [11] for references).

Interest in the topic increased sharply in the mid-90s. Generation, application and study of unipolar pulses has become extremely relevant with the advent of the era of few laser fields [13,14]. The main findings of [12] were again analyzed and confirmed [15,16]. The prospects and applications of unipolar pulses in microwave and optical ranges are presently being studied and reported in various new fields of science and technology. The parameter $S_E(r)$ is now also used in a broader sense than the criteria given by Bessonov formulae (1-3). For characterization of bipolar strange waves Arkhipov et. al [17] introduced the degree of unipolarity:

$$\xi(r) = \frac{|S_E(r)|}{\int_{-\infty}^{+\infty}|E(r,t)|dt} = \frac{\left|\int_{-\infty}^{+\infty} E(r,t)dt\right|}{\int_{-\infty}^{+\infty}|E(r,t)|dt}, \quad (4)$$

The method to measure $|S_E(r)|$-function of EM pulse using observation of quantum transitions induced by pulse interaction with a two-level system was proposed in [11].

In this work, it will be shown that in the absence of sources, any EM pulse is usual in the sense of the Bessonov condition (2). In other words, condition (2) is the same inherent property of EM pulses in vacuum as the invariants of energy, momentum, angular momentum, spin [2,18], the number of quanta [19-21], as well as STC, which was discussed above.

## 2. Materials and Methods: some relations for the vector potential in k-space.

First, let's repeat that we are talking about impulses of a general form in free space. The only requirement is the finiteness of total energy.

As in the method of expansion of EM field in terms of field oscillators [22], it is convenient to use the Fourier transform. This allows, instead of the field strengths $E(r,t)$ and $H(r,t)$, satisfying the free space Maxwell equations:

$$\text{rot}E(r,t) = -\frac{1}{c}\frac{\partial H(r,t)}{\partial t}, \ \text{div}E(r,t) = 0,$$
$$\text{rot}H(r,t) = \frac{1}{c}\frac{\partial E(r,t)}{\partial t}, \ \text{div}H(r,t) = 0. \quad (5)$$

consider only vector potential $A(r,t)$, which satisfies the equations:

$$\begin{cases} \Delta A(r,t) = \frac{1}{c^2}\ddot{A}(r,t) & (6a) \\ \text{div} A(r,t) = 0 & (6b) \end{cases}$$

The where $c$ is the speed of light in vacuum. In this case, the vectors of the electric and magnetic fields are expressed in terms of **A** as follows:

$$E(r,t) = -\frac{1}{c}\dot{A}(r,t); \quad H(r,t) = \text{rot}\, A(r,t). \tag{7}$$

The transition to the Fourier space in equations (6) after some transformations allows us to present the EM field as a set of independent harmonic oscillators. This is part of the standard procedure for quantizing the EM field. Further, for the purposes of this work, it is sufficient to express the fields $E(r,t)$ and $H(r,t)$ in terms of the spatial harmonics of the vector potential $A(k,t)$, which, as is known, depend on time according to the harmonic law. As a result, the proof is reduced to the study of integrals of rapidly oscillating functions.

For the Fourier harmonics of the vector potential $A(k,t)$, from Maxwell's equations in the form (6), it is easy to obtain the equations:

$$\ddot{A}(k,t) + k^2 c^2 A(k,t) = 0, \tag{8}$$

$$A^*(k,t) = A(-k,t), \tag{9}$$

$$(k \cdot A(k,t)) = 0. \tag{10}$$

Expressions for $E(r,t)$ and $H(r,t)$ are easily found from relations (7):

$$E(r,t) = -\frac{1}{c}\int \dot{A}(k,t)e^{ikr}dk,\, H(r,t) = i\int (k \times A(k,t))e^{ikr}dk,\, A(r,t) = \int A(k,t)e^{ikr}dk. \tag{11}$$

The total energy $\mathcal{E}$ does not depend on time:

$$\mathcal{E} = \int \epsilon(r,t)dr,\ \epsilon(r,t) = \frac{E^2(r,t)+H^2(r,t)}{8\pi}. \tag{12}$$

In $k$-space, taking into account (11) and (9), (10), expression (12) takes the form:

$$\mathcal{E} = \pi^2 \int [|E(k,t)|^2 + |H(k,t)|^2]d\vec{k} = \frac{\pi^2}{c^2}\int \left[|\dot{A}(k,t)|^2 + k^2 c^2 |A(k,t)|^2\right]d\vec{k}. \tag{13}$$

We now use equation (8), from which it obviously follows that:

$$A(k,t) = A_-(k)e^{-ickt} + A_+(k)e^{+ickt}, \tag{14}$$

and, therefore, the total energy (13) is equal to:

$$\mathcal{E} = 2\pi^2 \int k^2 [|A_-(k)|^2 + |A_+(k)|^2]d\vec{k}. \tag{15}$$

The functions $A_-(k)$ and $A_+(k)$ are determined by the initial conditions for equation (8). For each of them, the transversality condition similar to (10) is satisfied, and, in addition, they are linked by the relation:

$$A_+(k) = A_-^*(-k).$$

Thus, formula (14) expresses the vector potential of an arbitrary EM pulse of finite energy in terms of the vector functions $A_-(k)$ and $A_+(k)$, satisfying the transversality condition and square integrable with weight $k^2$. In the next section, this will be used to prove relation (3).

### 3. Results: free EM pulses are not strange.

Along with the strangeness vector $S_E(r)$ (see (1)), consider the vector $S_E(r,T)$ so that:

$$S_E(r) = \lim_{T\to\infty} S_E(r,T) = \lim_{T\to\infty}\int_{-T}^{T} E(r,t)\,dt = -\frac{1}{c}\lim_{T\to\infty}[A(r,T) - A(r,-T)]. \tag{16}$$

For the first term in (16), taking into account the last of formulas (11), as well as formula (14), we obtain:

$$\boldsymbol{A}(\boldsymbol{r},T) = \int \boldsymbol{A}(\boldsymbol{k},T)e^{i\boldsymbol{k}\boldsymbol{r}}d\boldsymbol{k} = \int \boldsymbol{A}_-(\boldsymbol{k})e^{i\boldsymbol{k}\boldsymbol{r}-ickT}d\boldsymbol{k} + \int \boldsymbol{A}_+(\boldsymbol{k})e^{i\boldsymbol{k}\boldsymbol{r}+ickT}d\boldsymbol{k}. \quad (17)$$

According to the Riemann-Lebesgue theorem [23], both terms in (17) disappear as $T \to \infty$ for any $\boldsymbol{r}$. The condition for the applicability of this theorem is the convergence of the integrals:

$$\int_0^\infty k^2 dk \left| \int d\widehat{\Omega}\, \boldsymbol{A}_-(\boldsymbol{k})\, e^{i\boldsymbol{k}\boldsymbol{r}} \right| \text{ and } \int_0^\infty k^2 dk \left| \int d\widehat{\Omega}\, \boldsymbol{A}_+(\boldsymbol{k})\, e^{i\boldsymbol{k}\boldsymbol{r}} \right|, \quad (18)$$

where $d\widehat{\Omega}$ means the integral over the angles of the vector k. Taking into account the convergence of the energy integral (15), the requirement of convergence of the integrals (18) does not seem excessively strict.

The second term on the right-hand side of (16) also disappears for $T \to \infty$. Thus, for a fairly large class of pulses in free space, we get:

$$\boldsymbol{S}_E(\boldsymbol{r}) = \lim_{T \to \infty} \int_{-T}^T \boldsymbol{E}(\boldsymbol{r},t)\, dt = 0, \quad (19)$$

cosequently relation (3) is satisfied. Hence, we can summarize this section by saying that each of the field projections of the free EM pulse of a finite energy is a sign-variable function of time, the integral of which is zero at any point in space.

## 4. Conclusions and discussion.

So, the answer to the question in the title of the article is: no. Free EM pulses are inevitably bipolar. Moreover, they satisfy Bessonov relation (3), i.e. their S-function (1) equals to zero everywhere. According to the duality principle (see [24]), this relation is also valid for the magnetic field $\boldsymbol{H}(\boldsymbol{r},t)$. The validity of what has been said is restricted by the conditions that determine the applicability of the Riemann-Lebesgue theorem for integrals of a vector potential (17).

Table 1 shows the classification of EM waves in terms of relations (2), (3), as well as the terminology introduced in section 1. Table 1 yields the conclusion: rigorous modeling of unipolar (single-sign) EM pulses and their applications requires accurate consideration of both the nature of the source and the propagation of the wave, since both factors affect the time shape of the pulse at the point of interest. A theoretical analysis of the radiation from a system of bound charges [12] and a laser medium bounded by a resonator [14] did not reveal unipolar waves.

In conclusion, we note the coexistence of two properties inherent in EM pulses of finite energy. On the one hand, this is the extreme variability in space and time associated with STC, and on the other hand, the equality to zero of the S-function in the whole space proved here. The latter can be used to control the accuracy of EM and QED computations with finite energy pulses, compare with [25, 26].

**Table 1.**

|  | EM pulses from moving charges [12] | | | Free space EM pulses (this paper) |
|---|---|---|---|---|
| charge motion | bounded | unbounded | | |
| $\boldsymbol{S}_E(\boldsymbol{r})$ | 0 | $\neq 0$ | 0 | 0 |
| usual or strange | usual | strange | usual | usual |
| bipolar or unipolar | bipolar | both are possible | bipolar | bipolar |

**Acknowledgments:** The authors are indebted to I.A. Artyukov, S.G. Bochkarev, V.Yu. Bychenkov, N.V. Dyachkov, R.M. Feshchenko, A.P. Kiselev, A.B. Plachenov and I.V. Smetanin for fruitfull discussions.